\documentclass[conference,10pt]{IEEEtran}
\IEEEoverridecommandlockouts
\usepackage{cite}
\usepackage{amsmath,amssymb,amsfonts}
\usepackage{algorithmic}
\usepackage{graphicx}
\usepackage{textcomp}
\usepackage{xcolor}
\usepackage{url}
\usepackage{amsmath,amsthm,amsfonts,amssymb,amscd}
\usepackage[ruled,vlined,linesnumbered]{algorithm2e}
\usepackage{commath}
\usepackage{subcaption}
\usepackage{float}
\usepackage{multicol}
\usepackage{colortbl}
\usepackage{multirow, makecell}

\def\BibTeX{{\rm B\kern-.05em{\sc i\kern-.025em b}\kern-.08em
    T\kern-.1667em\lower.7ex\hbox{E}\kern-.125emX}}
\begin{document}

\title{Open Set RF Fingerprinting using Generative Outlier Augmentation}

\author{\IEEEauthorblockN{Samurdhi Karunaratne, Samer Hanna, and Danijela Cabric 
\thanks{This work was supported in part by the CONIX Research Center, one of six centers in JUMP, a Semiconductor Research Corporation (SRC) program sponsored by DARPA.}
		}
		
\IEEEauthorblockA{\textit{Electrical and Computer Engineering Department,} \\
\textit{University of California, Los Angeles}\\
samurdhi@ucla.edu, samerhanna@ucla.edu, danijela@ee.ucla.edu }
}

\maketitle

\begin{abstract}
RF devices can be identified by unique imperfections embedded in the signals they transmit called RF fingerprints. The closed set classification of such devices, where the identification must be made among an authorized set of transmitters, has been well explored. However, the much more difficult problem of open set classification, where the classifier needs to reject unauthorized transmitters while recognizing authorized transmitters, has only been recently visited. So far, efforts at open set classification have largely relied on the utilization of signal samples captured from a known set of unauthorized transmitters to aid the classifier learn unauthorized transmitter fingerprints. Since acquiring new transmitters to use as known transmitters is highly expensive, we propose to use generative deep learning methods to emulate unauthorized signal samples for the augmentation of training datasets. We develop two different data augmentation techniques, one that exploits a limited number of known unauthorized transmitters and the other that does not require any unauthorized transmitters. Experiments conducted on a dataset captured from a WiFi testbed indicate that data augmentation allows for significant increases in open set classification accuracy, especially when the authorized set is small.
\end{abstract}


\section{Introduction}

With billions of new Internet of Things (IoT) devices added each year, the task of securing IoT networks has become more important by the day. However, traditional cryptography based authentication systems are typically not suited for authenticating such devices due to their limited power and computational constraints. To address this, passive Physical Layer Authentication (PLA) has been proposed since it requires little to no work on the part of the transmitter\cite{wang_wireless_2016}. Here, the authenticator uses channel state information and fingerprints due to hardware impairments to identify transmitters. Recently, research on passive PLA that uses deep learning techniques has been gaining momentum. Since deep learning based classifiers tend to extract more salient features, these approaches have been shown to outperform others which use handcrafted features, reaching markedly higher accuracy \cite{riyaz_deep_2018}. 

Most such authentication systems have been modeled as closed set classifications, where the classification decision is one of a known set of authorized transmitters. However, it is obvious that such a model is ill-suited for an authentication system which needs to discriminate among the authorized set while rejecting unauthorized transmitters---since any unauthorized transmitter not included in training would be misclassified as authorized. Therefore, open set classification, which can detect unseen transmitters, has recently been proposed in this regard \cite{gritsenko_finding_2019} \cite{hanna_open_2021}.

There are two primary methods in which openset classification problems can be solved using deep learning \cite{geng_recent_2019}. The first one is discriminative model-based methods. Here, authorized data (signal samples from authorized transmitters) and known unauthorized data (signal samples from some known unauthorized transmitters) may be used to train a specialized neural network architecture. Subsequently an accompanying detection method allows for the detection of outliers while detecting authorized classes. To this end, we have previously tested architectures like One vs All (OvA) and methods like OpenMax and demonstrated that, for example, using a higher number of known unauthorized transmitters during training leads to better performance \cite{hanna_open_2021}. However, in this paper, our focus is on the second approach to open set classification: instance-generation based methods \cite{geng_recent_2019}. As the name implies, here generative models are used to synthetically create samples from the unknown unauthorized data that could be used to augment the training dataset. Then discriminative model-based open set methods or even closed set methods could be used on the augmented training data. 

Data augmentation within the umbrella of RF fingerprinting is not an entirely new concept---for example, in \cite{soltani_more_2020} data with many simulated channel and noise variations were augmented to train a more channel-resilient RF fingerprinting classifier. However, to the best of our knowledge, instance-generation based data augmentation has not been applied in the literature regarding RF authentication. Yet generative models and methods have been proposed for data augmentation in other domains \cite{geng_recent_2019}. Inspired by this past work, we adapt some of these methods to RF fingerprinting while presenting our own methods. Our contributions include two types of outlier generation based on the type of information available during training:

\begin{enumerate}
    \item Supervised outlier generation: Some signal samples from a limited number of known unauthorized transmitters are available during training.

    \item “Blind” outlier generation: Only authorized signal samples are available during training.
\end{enumerate}
In both cases, the goal is to emulate signal samples from the set of unauthorized transmitters.

The rest of the paper is organized as follows: we start by formulating the problem in Section II. Section III discuses the dataset and evaluation method used for approaches presented throughout the paper. Supervised outlier generation is presented in Section IV while Blind outlier generation is presented in Section V. Section VI concludes the paper.

\providecommand{\mA}{\mathcal{A} }
\providecommand{\mK}{\mathcal{K} }
\providecommand{\mO}{\mathcal{O} }
\providecommand{\mX}{\mathcal{X} }
\providecommand{\mY}{\mathcal{Y} }
\providecommand{\mYh}{\mathcal{\hat{Y}} }

\providecommand{\mAc}{\mathcal{|A|} }
\providecommand{\mKc}{\mathcal{|K|} }
\providecommand{\mOc}{\mathcal{|O|} }
\providecommand{\mYhc}{\mathcal{|\hat{Y}|} }

\section{System Model and Problem Formulation}

\begin{figure}[t]
    \begin{center}
    \vspace*{18px}
    \includegraphics[width=0.49\textwidth]{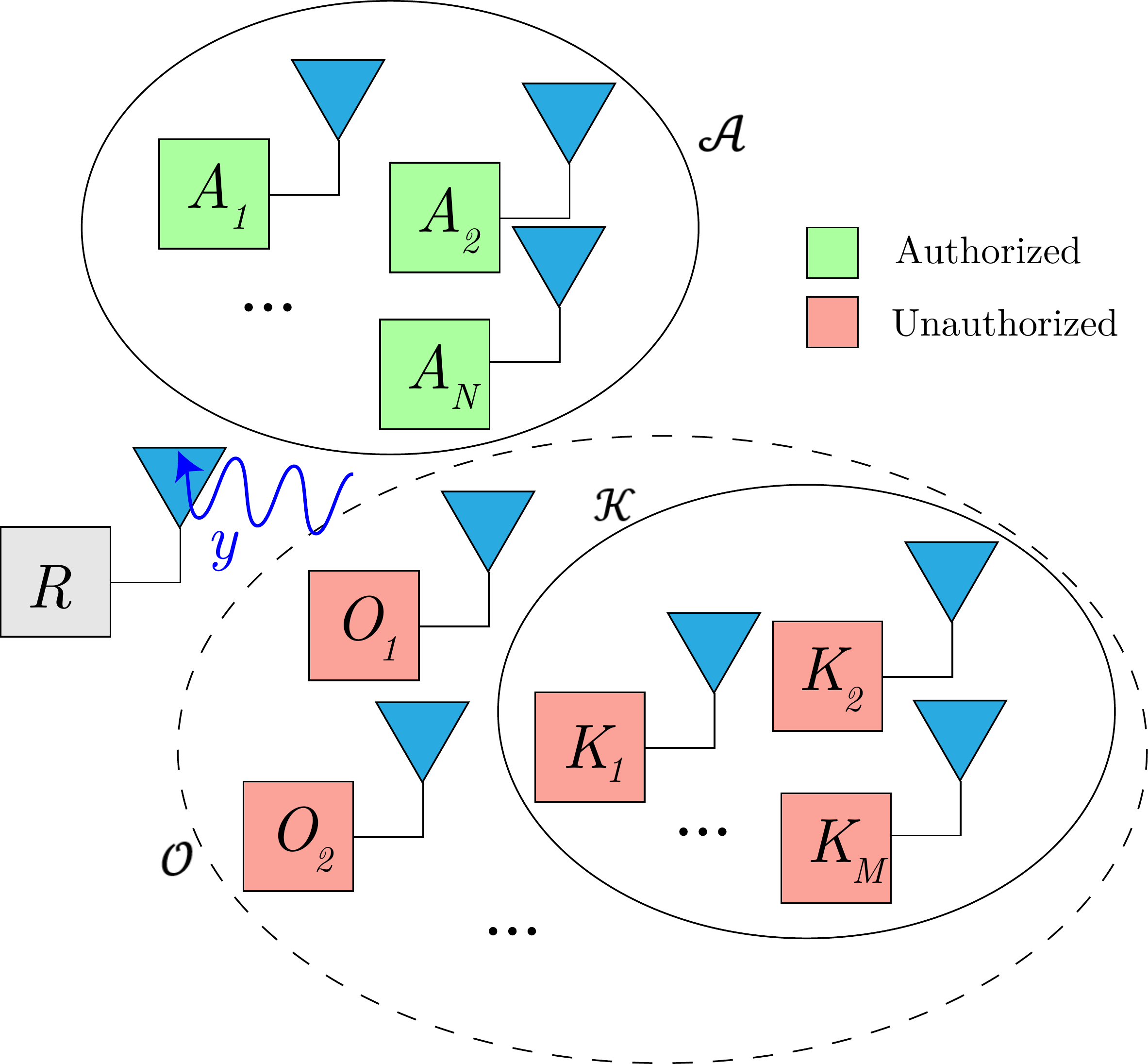}
    \end{center}
    \setlength{\belowcaptionskip}{-5pt} 
    \captionsetup{belowskip=-5pt}
    \caption{System model: $R$ must determine whether the received signal $y$ originated from an unauthorized transmitter in $\mO$, and if not, from which one of the authorized transmitters in $\mA$. Some unauthorized transmitters $\mK$ may be known to $R$.}%
    \label{fig:system_model}%
\end{figure}

We consider a finite set of authorized transmitters given by  $\mA=\{A_1,A_2, \cdots,A_\mAc \}$ that are authorized to send data to a receiver~$R$, where $\mAc$ is the size of the set $\mA$. When a transmitter $T$ sends a set of symbols $x$, the signal received is $f_T(x)$. The function $f_T$ models the transmitter fingerprint determined by the variability of its circuits and also includes the effects of the  channel. The authentication problem can be formulated as shown in Fig.~\ref{fig:system_model}: receiver $R$ receives a signal $y$ from some transmitter $T$ and should determine whether the transmitter $T$ belongs to the authorized set or not, based on $y$. This can be formulated as the following hypothesis testing: 
\providecommand{\mPfa}{P_{FA}}
\providecommand{\mPd}{P_{D}}
\newcommand{\mHz}{\ensuremath{\mathcal{H}_0} }
\newcommand{\mHo}{\ensuremath{\mathcal{H}_1} }
\renewcommand{\b}[1]{\boldsymbol{\mathrm{#1}}}
\begin{align}
\begin{split}
	\mHz: & \ y = f_T(x), T\in \mA \\
	\mHo: & \ y = f_T(x), T\notin \mA  
\end{split}
\end{align}  

However, in this paper, we are additionally interested in classifying the transmitter within the authorized set, in cases where each authorized transmitter has different privileges. In such a setting, the open set authorization problem could be written as
\begin{align}
\begin{split}
	\mathcal{H}_0: & \ y = f_T(x), T = A_1 \\
	\mathcal{H}_1: & \ y = f_T(x), T = A_2 \\
	\cdots \\
	\mathcal{H}_{\mAc -1}: & \ y = f_T(x), T = A_{\mAc} \\
	\mathcal{H}_{\mAc}: & \ y = f_T(x), T\notin \mA
\end{split}
\label{eq:hypotheses_problem}
\end{align} 
To improve the outlier detection, an additional set of known outliers $\mK=\{K_1,K_2,\cdots,K_\mKc\}$, where $\mK \not\subset \mA$, may be used. So typically during training, a captured set of signal samples $\mX=\{X_1,X_2,\cdots,X_\mAc\}$ corresponding to transmitters in $\mA$ and a similar set of signal samples $\mY=\{Y_1,Y_2,\cdots,Y_\mKc\}$ corresponding to transmitters in $\mK$ will be used during training to assist the outlier detector to differentiate between authorized and non-authorized transmitters.
In practice, samples from the set $\mK$ can be obtained by capturing data from a finite number of non-authorized transmitters. 

During supervised outlier generation, the task is to use the information in $\mY$ (and possibly $\mX$) to generate more signal samples from $\mO$, which is the theoretically infinite set of unauthorized transmitters ($\mK \subset \mO$). The task in blind outlier generation is to use only the information encoded in $\mX$ to generate more signal samples from $\mO$. The set of such samples generated from $\mO$, using either method, is denoted by $\mYh$.

\section{Dataset and Evaluation Method}

We introduce the dataset and outlier detector classifer architecture for evaluating the proposed approaches.

 \subsection{Dataset}
 The dataset was captured on the Orbit testbed \cite{orbit_2005}---each one of the 71 transmitters was one 3 off-the-shelf WiFi modules (Atheros 5212, 9220, and 9280) and the receiver was a software defined radio (USRP N210).

Each transmitter was allowed to transmit over Channel 11, which has a center frequency of 2462 MHz and a bandwidth of 20 MHz, and captures were taken at a rate of 25 Msps for one second. After the IQ capture was complete, energy detection was used to detect the start and end of packets and extract them. Due to WiFi rate control, the number of packets obtained from each transmitter varied between 200 and 1500 packets with a mean of 800 packets. From each packet, we used the first 256 IQ samples, containing the preamble, without any synchronization or further preprocessing, as the signal sample.

\subsection{Outlier detector architecture and evaluation}

\begin{figure}
    \centering
    \includegraphics[width=0.8\linewidth]{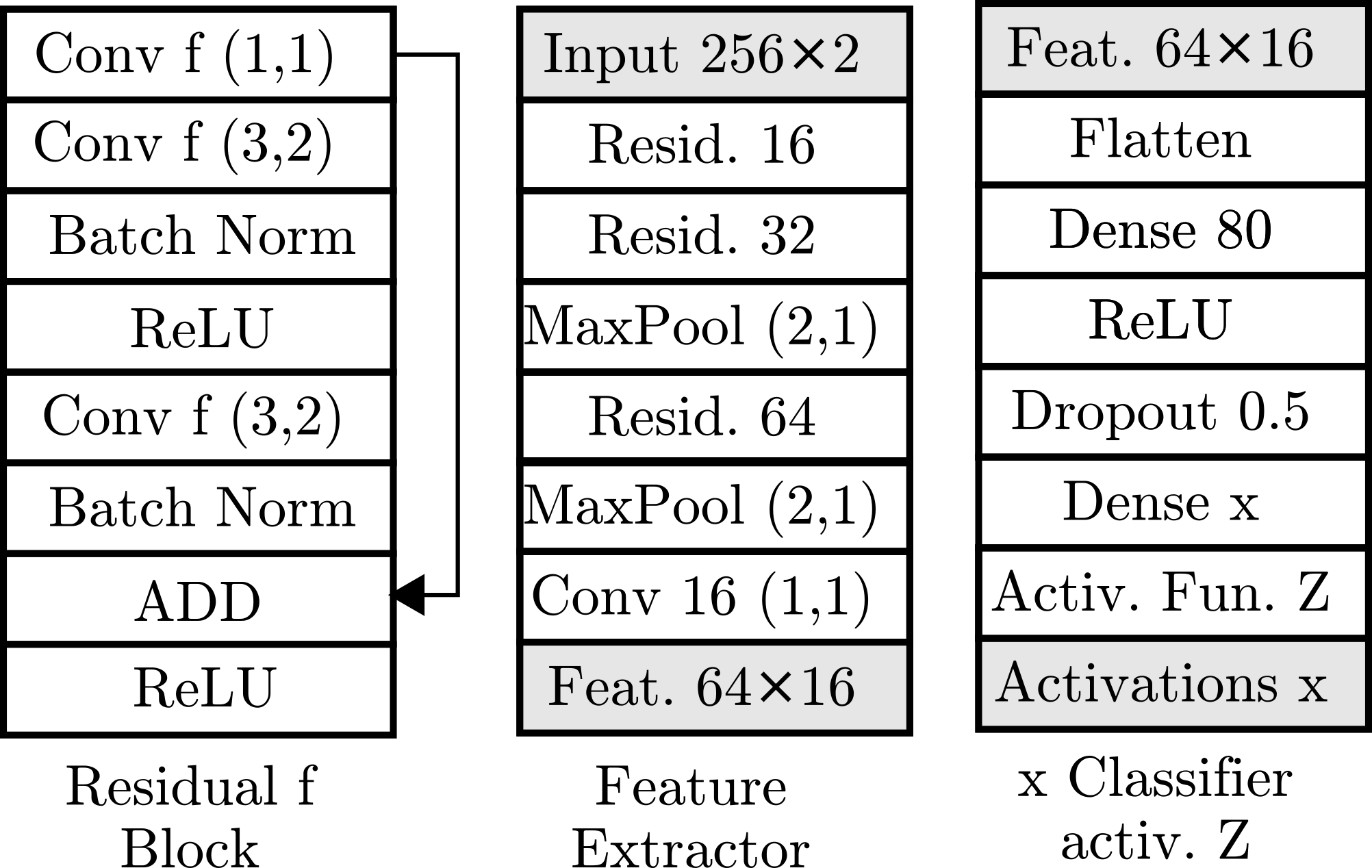}
    \caption{Detailed architecture of the feature extractor (made of residual blocks with f filters), and a classifier with x outputs.}
    \label{fig:classifier_layers}
\end{figure}

In \cite{hanna_spawc_2020},  we explored several neural network architectures that could be used for a problem like Eq. \ref{eq:hypotheses_problem} such as DClass and OvA. In this paper, for our evaluations, we chose to use the OvA (One-versus-All) architecture, most importantly because it could be used even when there are no known outliers during training. Simply put, the OvA architecture consists of a feature extractor followed by $\mAc$ binary classifiers in parallel, where the $i$-th binary classifier decides (0 or 1) whether the input signal is from $A_i$ or not. The feature extractor architecture used for all experiments conducted in this paper is given in Fig. \ref{fig:classifier_layers}.

$\mA$, $\mK$ and $\mO$ will be chosen randomly, subject to the constraints specified for each evaluation---however, when comparing a non-augmented dataset vs. the corresponding augmented dataset, the same $\mA$, $\mK$ and $\mO$ will be kept. For chosen $\mA$, $\mK$ and $\mO$, the dataset split will be as follows: for training and validation, we use 70\% of the samples belonging to $\mA$, and all the samples belonging $\mK$. The shuffled combination of this data is split into 80\% for training and 20\% for validation. The
test set contains all samples from $\mO$ and the remaining 30\%
of $\mA$. The evaluation metric is the prediction accuracy of the OvA classifier on this test set.

\section{Supervised outlier generation}

The idea behind supervised outlier generation is simple: we could train a generative model on signal samples collected from a few known outliers ($\mY$), and use that generative model to generate more outlier samples. We chose Variational autoencoders (VAEs) \cite{kingma_auto-encoding_2014} as our preferred generative model due to their relative ease of training. A VAE is an unsupervised neural network architecture that consists of an encoder network $E$ which transforms the input data into a latent space and a decoder network $D$ which maps the encoded data back into the input space. By learning parameters related to the underlying data distribution, for any standard normal $z$, it is able to take $D(z)$ as a sample from that distribution.

\begin{figure}
    \centering
    \includegraphics[width=0.9\linewidth]{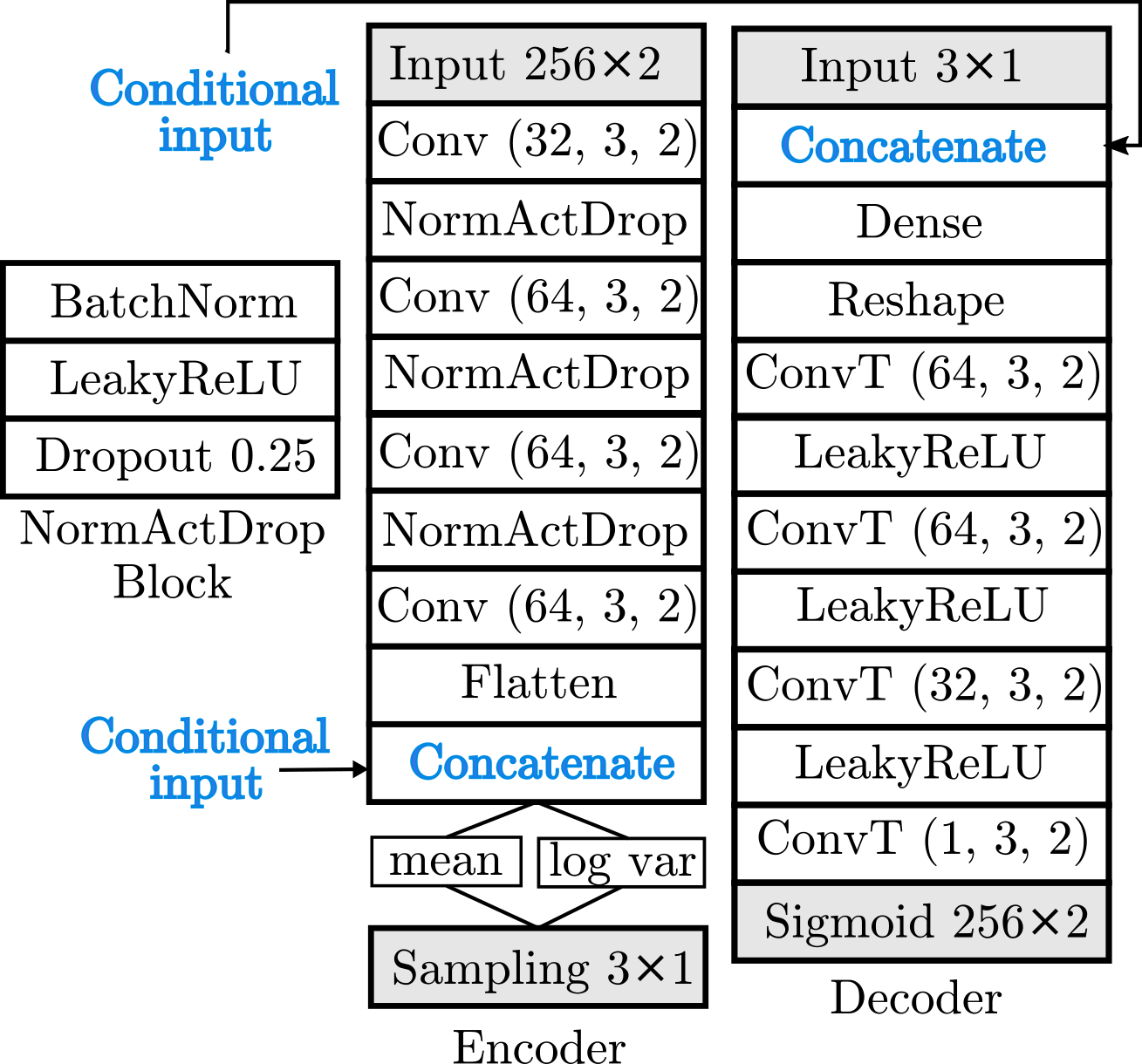}
    \caption{Neural network architectures used for the standard VAE and the conditional VAE. The components in \textcolor{blue}{blue} are only present in the conditional VAE. \texttt{Conv(x,y,z)} is a 2D Convolutional layer with $x$ filters of kernel size $y$ and stride ($z$, 1). \texttt{ConvT} represents a transposed version of \texttt{Conv}.}.
    \label{fig:vae_architecture}
\end{figure}

\begin{figure}[t]
    \subfloat[{Variation of testing accuracy with $\mKc$ for VAE}]{\label{fig:vae_performance}\includegraphics[width=1\linewidth]{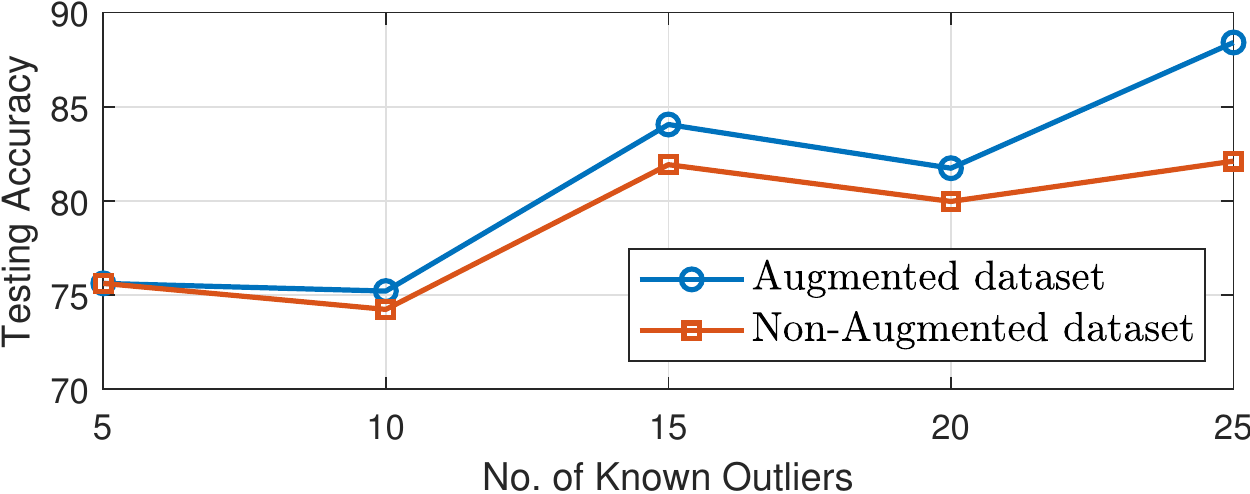}}
    
    \subfloat[{Variation of testing accuracy with $\mKc$ for CVAE}]{\label{fig:cvae_performance}\includegraphics[width=1\linewidth]{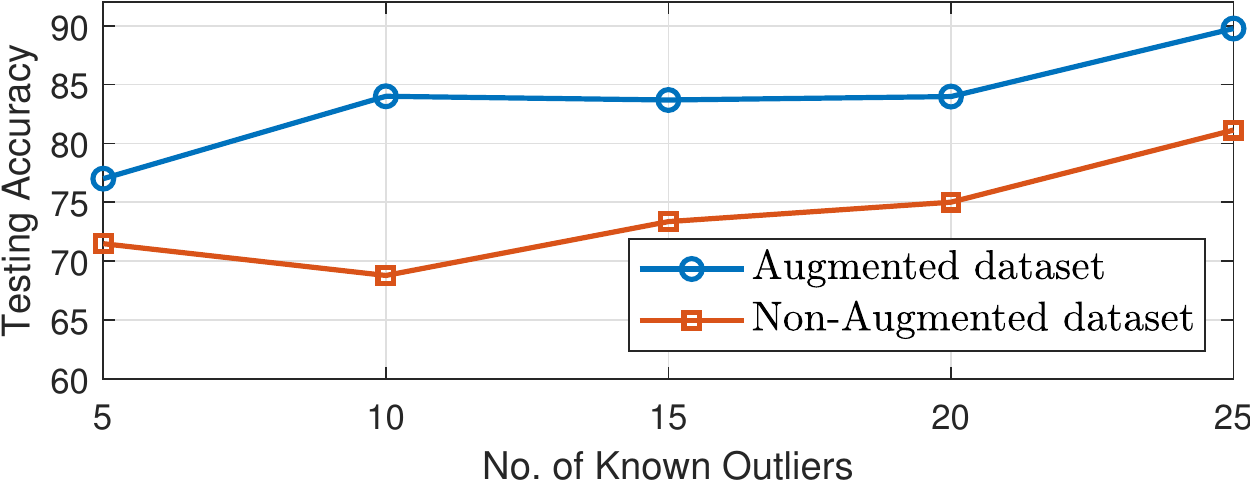}}
    \caption{Variation of testing accuracy with $\mKc$ for VAE and CVAE. }
    \label{fig:supervised_outlier_generation_results}%
\end{figure}

The most straightforward approach is to train a  VAE on all the data available in $\mY$. Intuitively, if we then isolate the decoder $D$, it should be possible to emulate as many known outlier samples as we want. To test this approach, we first created a  VAE with the architecture highlighted in Fig. \ref{fig:vae_architecture} and performed the following experiment: we pick a fixed set of 10 authorized transmitters $\mA$. Then on 5 different occasions, we pick $\mK$ randomly from $\mO$ such that $\mKc \in \{5, 10, 15, 20, 25\}$. These values for $\mAc$ and $\mKc$ were picked since we seek to evaluate the effect of $\mKc$ on outlier detection performance as $\mAc$ is kept constant---any value for $\mAc$ and any set of values for $\mKc$ not too close together should give similar observations for what follows. For each such $\mK$, a  VAE was trained on the corresponding $\mY$ and its decoder was then used to generate more samples from $\mK$. We then train an OvA classifier with the individually labelled $\mX$ as authorized and $\mY$ and $\mYh$ as the outliers. In our experiments it was interesting to note that while the testing accuracy of the trained OvA as $\mYhc$ is increased rises initially, beyond a certain number of generated samples, there is no appreciable increase. This makes sense because the finite set $\mK$ and the finite number of captured signal samples $\mY$ only expose a certain amount of information about $\mO$ that could be exploited by generating more samples from $\mK$. Therefore we fixed $\mYhc=7500$ and evaluated the OvA testing accuracy when the training data is augmented with the generated outliers and compared it against the non-augmented OvA performance. From Fig. \ref{fig:vae_performance} it is clear that when $\mKc$ is increased, the effect of augmentation is greater. While many reasons are possible, it is immediately not clear why this trend is observed---so we experiment further.

Note that when trying to generate samples from $\mK$ by training a standard VAE on $\mY$ as above, we do not know from which transmitters in $\mK$ these generated signals samples are drawn from. For example, if 1000 samples are generated and $\mKc = 5$, 800 of those could be from $K_1$ and 200 could be from $K_3$, or all 1000 could be from $K_4$. To avoid this disproportionality during the sample generation, we tested another type of VAE called a conditional VAE or CVAE. The CVAE is simply an extension on the standard VAE in which there is a conditional or categorical input to the decoder indicating the class of the sample to be generated, as is clear from Fig. \ref{fig:vae_architecture}. This allows us to specify the particular transmitter from which samples are to be generated after the CVAE is trained. Similar to the VAE, for each $\mKc \in \{5, 10, 15, 20, 25\}$ we generated 7500 samples but this time, $7500 / \mKc$ samples were generated from each known outlier in $\mK$. Then the same OvA classifier as for the standard VAE was trained---the testing accuracy observed for different $\mKc$ is given in Fig. \ref{fig:cvae_performance}. We can see that now roughly the same level of accuracy increase is seen for all $\mKc$---when $\mKc$ is large, the probability that generated samples will be picked from a small subset of $\mK$ is small, while that same probability is large when $\mKc$ is small. So using a CVAE has an equalizing effect by allowing generated samples to be proportionately chosen. This is why we see the augmentation accuracy gap increase with $\mKc$ for the standard VAE while roughly staying the same for the CVAE. Note that the non-augmented curves in Fig. \ref{fig:vae_performance} and Fig. \ref{fig:cvae_performance} are different due to the randomness in which $\mK$ is selected.


\section{Blind outlier generation}

While supervised outlier generation is somewhat straightforward, it requires the need to acquire additional hardware in the form of known unauthorized transmitters. This is an expensive task that becomes infeasible in many situations. So a more important, and more challenging task is to use just the information available in $\mX$ to emulate signal samples from $\mO$: blind outlier generation.

\subsection{Ellipsoisal method}

Our approach to this task is rooted in visualizing the authorized set in some latent space, which is essentially a dimensionality-reduced space. Any signal $x \in X$ is typically a high-dimensional vector ($256\times 2=512$-dimensional in the dataset above), so this dimensionality reduction allows for correlated features to be merged and the dimensionality of the vector space to be made tractable. Here, we will use an \emph{autoencoder} to map the authorized signal samples to a latent space (as will become evident, a VAE is unnecessary as we do not require any sampling in this case). Since the size of the encoder output is usually designed to be much smaller than the size of the input, the encoder essentially extracts a latent space representation of the input.

\begin{figure}[t]
    \centering
    \includegraphics[width=1\linewidth]{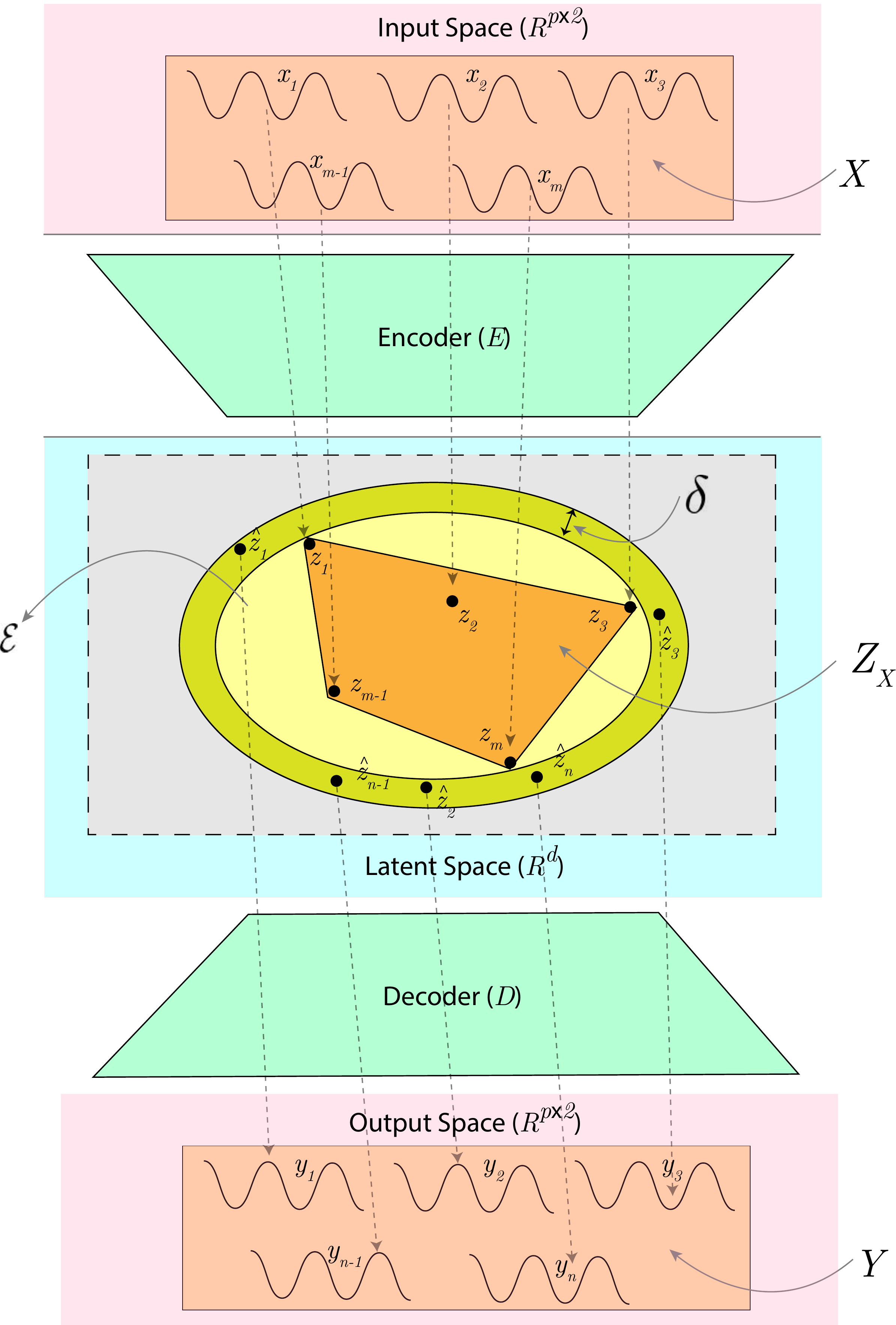}
    \caption{Ellipsoidal method: Using an encoder to map authorized samples $x_i$ to the latent space $z_i$. Random points sampled in the shell are mapped using the decoder to recover generated outlier samples $y_j$}
    \label{fig:ellipsoidal_method_schematic}
\end{figure}

This process is visually depicted by Fig. \ref{fig:ellipsoidal_method_schematic} --- we start off with some authorized signal samples $X=\{x_1, x_2, ..., x_m\}$ that we map to their corresponding latent space representations $Z_X=\{z_1, z_2, ..., z_m\}$ using the encoder, i.e. $z_i = E(x_i)$.
Note that the region in latent space occupied by $Z_X$ could be some arbitrary shape. The key insight behind our approach to blind outlier generation is to conjecture that in latent space: i) outliers are still wireless signals so they should be ``close" to authorized signal samples within the latent space; and 
ii) outliers are clearly not authorized signals so they should be clearly separated from the region of space occupied by the authorized signals.

Now assume we were able to find the minimum volume ellipsoid $\mathcal{E}$ enclosing $Z_X$. Then, outlier signals should be located in some region of the latent space that is separate from but not too far away from the region occupied by the authorized signals.
This suggests that if we take some shell of uniform thickness $\delta$ around $\mathcal{E}$, points within this region should correspond to outlier signals.

To generate outlier samples in the signal space, we could simply generate random points within the shell and use the decoder $D$ to map them into their signal space counterparts. This process is visualized in Fig. \ref{fig:ellipsoidal_method_schematic}. 

To practically apply this method, we first need a method to find the minimum volume ellipsoid $\mathcal{E}$. We can write 
\begin{align}
    \mathcal{E} =\{z \in Z_X ~|~ \Vert Az+b \Vert_2 \leq 1\}
\end{align}
i.e. $\mathcal{E}$ is the inverse image of the Euclidean unit ball under an affine mapping, and the variables are $A \in \mathbb{R}^{n \times n}$ and $b \in \mathbb{R}^{n}$ where $A$ could be assumed to be positive definite without loss of generality.

Since the volume of $\mathcal{E}$ is proportional to $\det A^{-1}$, this can be formulated as the following convex optimization problem, which can be solved numerically
\begin{align}
\begin{array}{lll}
\text{minimize}_{A\in S^n,b\in \mathcal{R}^n} & \log \det A^{-1} &\\
\text{subject to} & \Vert Az_i+b\Vert_2 \leq 1 & i=1,\dots,m
\end{array}
\end{align}

\begin{figure}[t]
    \subfloat[{Sweet spot for $\delta$: Variation of testing accuracy as thickness of the ellipsoid shell is tuned for $\mAc=5$ }]{\label{fig:ellipsoidal_sweet_spot}\includegraphics[width=1\linewidth]{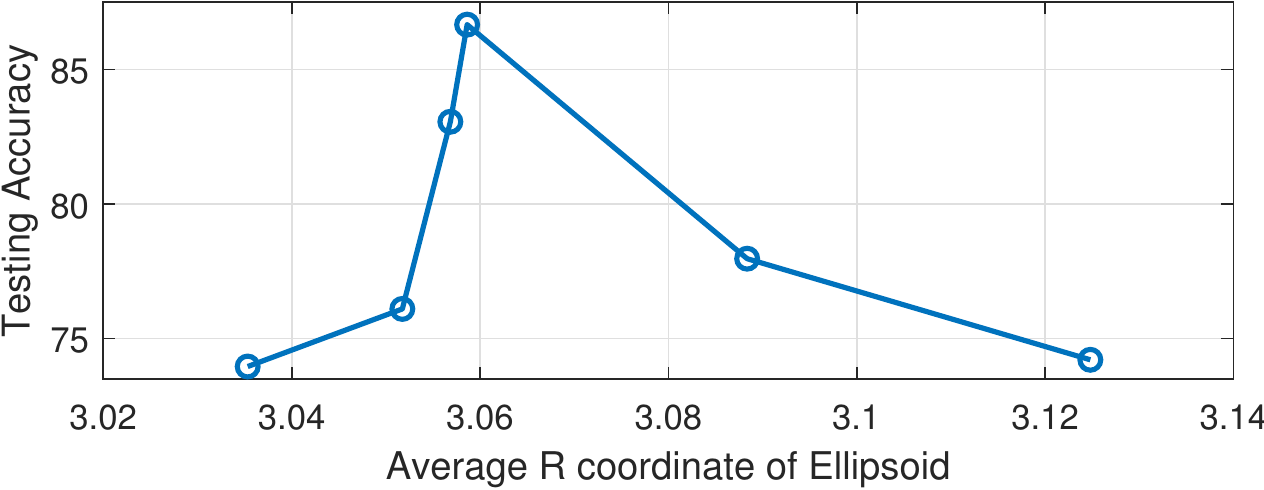}}
    
    \subfloat[{Variation of testing accuracy against no. of authorized transmitters}]{\label{fig:ellipsoidal_accuracy}\includegraphics[width=1\linewidth]{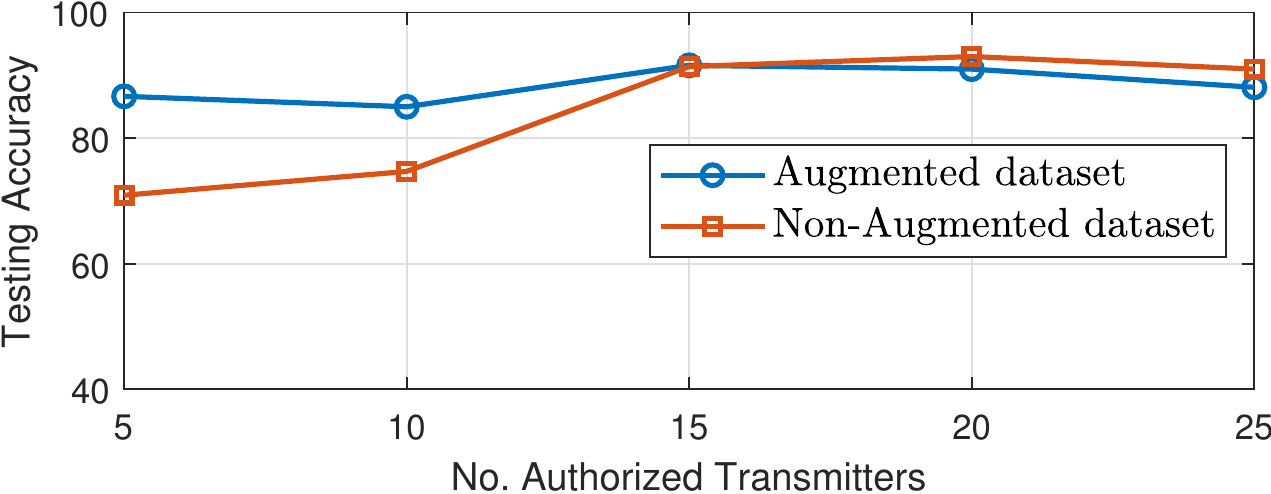}}
    \caption{{The results of using the ellipsoidal method to generate outliers}}%
    \label{fig:ellipsoidal}%
\end{figure}

To evaluate this method, we performed the following experiment. 30 transmitters were first selected from the 71-transmitters in the dataset as the set of outliers ($\mOc = 30$). Then for $\mAc \in \{5, 10,15,20,25\}$, a set of $\mAc$ transmitters were randomly selected as the authorized set $\mA$ in 5 separate occasions. Note that similar to the VAE methods in Section IV, even though it is technically possible to use this method to generate an infinite number of outliers, practically we see almost no improvement in performance after a certain number of outlier samples are generated. So as we did in Section IV, $\mYhc=7500$ outlier samples were generated and subsequently used to augment the training dataset and train an OvA classifier. The testing accuracy corresponding to outlier samples that were generated for different values of $\delta$ for $\mAc=5$ is given in Fig. \ref{fig:ellipsoidal_sweet_spot}. We can clearly see that there is an ideal value for the shell thickness $\delta$, that is not too small and not too large, when the generated samples represent the best outliers.  To generate Fig. \ref{fig:ellipsoidal_accuracy}, the same $\delta$ that was ideal for $\mAc = 5$  was used for all $\mAc $ when generating $\mYh$. In Fig. \ref{fig:ellipsoidal_accuracy}, we can see that especially when $\mAc$ is low, significant gains ($>$15\%) in testing accuracy could be obtained by augmenting the dataset.

\subsection{Optimization method}

The ellipsoidal method provided a visually intuitive method for outlier generation. However, it required the empirical tuning of a parameter $\delta$. So we try to approach the same problem with a similar intuition about outlier signals but now formulate the constraints of the problem differently. As we had previously remarked, the outlier signals should correspond to wireless signals. So if $D(z)$ is an outlier, the encoding $E(x)$ of an authorized signal $x$ should be close to $z$ in latent space. For $D(z)$ to be an outlier, $z$ should satisfy,
\begin{align}
\begin{array}{ll}
\text{minimize}_{z} & \Vert E(x)-z\Vert\\
\text{subject to} & D(z) \notin \mathcal{A}
\end{array}
\label{eq:optimization_formulation}
\end{align}
    Here, $||\cdot||$ could be some norm measure like the Euclidean distance. However the constraint $D(z) \notin \mathcal{A}$ is less straightforward. To solve this problem and find some optimal $z$ for some input $x$, we need to integrate this constraint numerically. Authors in \cite{ferrari_open_2018} suggested using a closed set classifier $C_{\mA}$ with $|\mA|$ outputs, which has learned to differentiate the known classes, as a judge of whether $D(z)$ is an outlier or not. They argued that if $D(z)$ is an outlier, the logits of the classifier prediction $C_{\mA}(D(z))_i$, should be low for all classes $i$. Inspired by this, we suggest using a closed set classifier $C_{\mA+1}$ with $|\mA|+1$ outputs as the judge of whether $D(z)$ is an outlier instead. With this definition, an ideal outlier would result in a softmax output of $\begin{bmatrix}0&0&\dots&0&1\end{bmatrix}$ meaning that we can formulate the constraint in terms of a loss function and integrate in to the objective function:
\begin{equation}
z^{*} = \min_z \Vert E(x)-z\Vert+ \text{loss} \left( \begin{bmatrix}0&\dots&0&1\end{bmatrix}, C_{\mA+1}(D(z)) \right)
\label{eq:optimization_obj_func}
\end{equation}
Using an $|\mA|+1$ output classifier as the judge instead of an $|\mA|$ output classifier has one added benefit---once we have found some set of initial outliers $\mO^{(0)}$, we can re-train the initial judge $C_{\mA+1}^{(0)}$ using those outliers as unauthorized samples to get a better judge $C_{\mA+1}^{(1)}$ that could be used in a subsequent iteration to generate even better outliers. Note that during its initial training $C_{\mA+1}$ does not have training samples for the $|\mA|+1$-th class, so this process should help it become a better judge as iterations progress. However, due to the self-supervised nature of this procedure, it could be sensitive to the quality of initial training of $C_{\mA+1}^{(0)}$. This method is highlighted in Algorithm 1 in detail.

\begin{algorithm}
\begin{small}
\caption{Blind outlier generation}
\SetAlgoLined
\SetKwInOut{Input}{Input}
\SetKwInOut{Output}{Output}
\SetKwInOut{Initialization}{Initialization}
\Input{Set of authorized signals $X$\;
}
\Output  {Samples from the set of outliers    $\mathcal{O}$\;}
Train encoder network $E(x)$, decoder network $D(z)$ and $\mAc+1$ closed set classifier $C_{\mA+1}^{(0)}$\;
\For {$i \in \{0,1,\dots,N-1\}$}
{
\textbf{initialize} $\mathcal{O}^{(i)} = \emptyset$\;
\For {$x \in X$}
 {
  $z^{*} = \min_z \Vert z-E(x)\Vert+ \text{loss} \left( \begin{bmatrix}0&0&\dots&0&1\end{bmatrix}, C_{\mA+1}^{(i)}(D(z)) \right)$\;
$\mathcal{O}^{(i)}=\mathcal{O}^{(i)} \cup \{D(z*)\}$
 }
$C_{\mA+1}^{(i+1)}=\text{train}(C_{\mA+1}^{(i)},\mathcal{X} \cup \mathcal{O}^{(i)})$\;
}
$\mathcal{O}=\mathcal{O}^{(N-1)}$\;
\end{small}
\end{algorithm}

\begin{figure}
    \centering
    \includegraphics[width=1\linewidth]{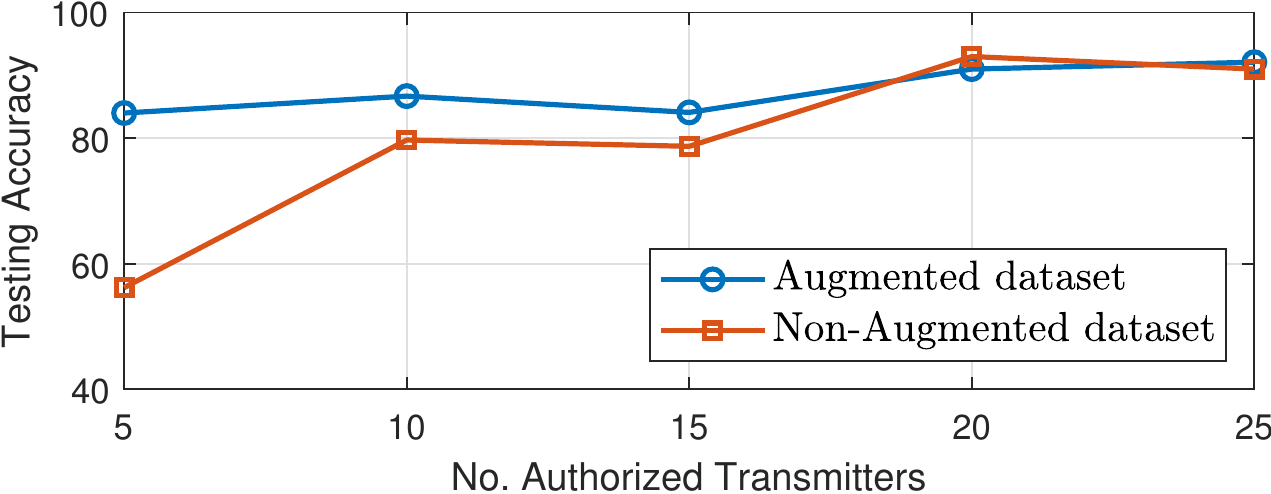}
    \caption{Variation of testing accuracy against no. of authorized transmitters for the optimization method}
    \label{fig:optimization_results}
\end{figure}

To evaluate this approach, we follow a similar approach to the ellipsoidal method. The architectures for the $E$ and $D$ networks are the same as that in Fig. \ref{fig:vae_architecture}, while cross-entropy was used as the loss function in Eq. \ref{eq:optimization_obj_func}. The results in Fig. \ref{fig:optimization_results} are almost identical to that of the ellipsoidal method in Fig. \ref{fig:ellipsoidal_accuracy}, except that we are able to see an improvement in testing accuracy even at $\mAc=15$, which was not seen with the ellipsoidal method. The crucial benefit is that there is no tuning of an extra hyperparamter ($\delta$) necessary in this case. However, the ellipsoidal method is still a valid method to consider if proper tuning of $\delta$ can be done, as solving Eq. \ref{eq:optimization_obj_func} through gradient descent is computationally expensive. For example, generating $10^3$ outlier samples with the optimization method took $30\times$ the time it took for the ellipsoidal method. Note that the non-augmented curves in Fig. \ref{fig:ellipsoidal_accuracy} and Fig. \ref{fig:optimization_results} are slightly different due to the randomness in which $\mA$ is selected.

One reason for the degradation in the augmentation accuracy gap in  Fig. \ref{fig:optimization_results} as $\mAc$ increases is that the outlier detection accuracy of $C_{\mA+1}$ decreases as $\mAc$ increases, due to there being no supervision for the outliers during the initial training of $C_{\mA+1}$ (it is an $\mAc + 1$-class classifier). So $C_{\mA+1}$ becomes worse at judging the constraint $D(z) \notin \mathcal{A}$ in Eq. \ref{eq:optimization_formulation}, affecting the quality of the outliers thereby produced.

\section{Conclusion}

In this paper, we considered the instance-generation based method for the open set classification problem in RF fingerprinting. Outlier instances were generated in two primary settings according to the information used during training; supervised outlier generation, where signal samples from a limited number of known outliers are used to generate more outlier samples; and blind outlier generation where outlier samples must be generated by using only authorized samples available during training. For supervised outlier generation, we considered VAEs and CVAEs and showed that CVAEs allows for a more uniform performance improvement when $\mKc$ is varied by enabling the proportionate generation of outliers from each known outlier in $\mK$. For blind outlier generation we first presented the ellipsoidal method, which was a geometrically intuitive way. Although it gave satisfactory results, it required the tuning of an extra hyperparameter. So another optimization-based method was presented that achieved the same-level of results but did not require any tuning. In both cases, significant performance improvements were seen for small $\mAc$. This provides evidence that generative outlier augmentation can replace the need to acquire transmitters as known outliers when building RF authentication systems.

\bibliographystyle{ieeetr}

\end{document}